\documentstyle[prc,aps]{revtex}
\begin{document}
\draft
\title {Cross section of $^{36}$S(n,$\gamma$)$^{37}$S}
\author{H.~Beer}
\address{Forschungszentrum Karlsruhe, Institut f\"ur Kernphysik,
P.~O.~Box 3640, D--76021 Karlsruhe, Germany}
\author{P.~V.~Sedyshev, Yu.~P.~Popov}
\address{Frank Laboratory of Neutron Physics, JINR,
141980 Dubna, Moscow Region, Russia}
\author{W.~Balogh, H.~Herndl, H.~Oberhummer}
\address{Institut f\"ur Kernphysik, Wiedner Hauptstr.~8--10, TU Wien,
A--1040 Vienna, Austria}
\date{\today}
\maketitle
\begin{abstract}
At the Karlsruhe pulsed 3.75\,MV Van de Graaff accelerator the
$^{36}$S(n,$\gamma$)$^{37}$S(5.05\,min) cross section was measured by the
fast cyclic
activation technique via the 3.103\,MeV $\gamma$--ray line of the
$^{37}$S--decay.
Samples of elemental sulfur enriched in $^{36}$S by 5.933\,\%
were irradiated between two gold foils which served as capture standards.
The capture cross section was measured at the neutron energies 25, 151,
176, and 218\,keV, respectively.
The $^{36}$S(n,$\gamma$)$^{37}$S--cross section in the thermonuclear
and thermal energy range has been calculated using the direct--capture
(DC) model
combined with the folding procedure used for the determination of the
potentials. The non--resonant experimental
data for this reaction can be reproduced excellently using this
method. The input parameters of the DC--calculation
(masses, Q--values, nuclear density distributions,
spectroscopic factors, spin--parity assignments and excitation energies
of the low--lying states of the residual nucleus) have been taken
from the available experimental data.
\end{abstract}


% 25.40.Lw: Radiative capture
% 25.40.Ny: Resonance reactions
% 25.40.Dn: Elastic neutron scattering

\section{Introduction}\label{s1}
In the last years the importance of the direct--reaction
(DI) mechanism in nucleosynthesis has been realized.
The DI dominates over the
compound--nucleus (CN) reaction mechanism if there exist no
CN--levels near the threshold that can be excited in the
reaction. For instance, this can be the case in
reactions involving light nuclei,
in the  big--bang scenario
as well as in in stellar hydrogen and helium burning. Direct capture (DC)
can also be of importance in proton capture by proton--rich target
nuclei~\cite{har88}--\cite{her95}
in the rp--process occurring in novae or X--ray bursts and
in neutron capture by neutron--rich nuclei~\cite{bee92}--\cite{mei95}
in the inhomogenous big bang as well as in the s--process taking place
in helium burning of red giants.
Furthermore, DC can dominate for neutron capture by light~\cite{gru94}
and heavy~\cite{mat83}--\cite{rau95} target nuclei far from
stability in the $\alpha$--process
and r--process occurring in supernovae of type II, respectively.
The reaction rates of the neutron--rich S--isotopes are of
interest in the nucleosynthesis of nuclei in the s--process
in the S--Cl--Ar--Ca region~\cite{bee87}, \cite{dru94},
inhomogenous big--bang scenario~\cite{mal93}, \cite{rau94},
and in the $\alpha$--rich freeze out of the
neutron--rich hot neutrino bubble in supernovae of
type II~\cite{woo92}--\cite{wit94}.

For the rare isotope $^{36}$S a significant abundance contribution is
expected from the s--process nucleosynthesis. For quantitative
analyses the size of the destruction rate, i.e.~the neutron capture rate,
is of fundamental importance
to estimate the magnitude of the $^{36}$S abundance formed by the weak and
main s--process components.
Using the statistical model the $^{36}$S capture cross
section has been estimated to be 300\,$\mu$barn at 30\,keV by
Woosley et al.~\cite{woo78}. The present measurement, the first
experimental
investigation of this cross section, applies the fast cyclic activation
technique~\cite{bee94} developed at the Karlsruhe
3.75\,MV Van de Graaff accelerator.

We investigate the capture reaction $^{36}$S(n,$\gamma$)$^{37}$S
from thermal (25.3\,meV) to thermonuclear (25\,--218\,keV)
projectile energies
and compare the calculated cross sections in the DC--model
with the experimental data.
In Sections~\ref{s2} and~\ref{s3} the activation technique is
described, and the experimental results are given.
In Section~\ref{s4} we introduce the methods used for calculating the
direct--capture (DC) cross sections.
In Section~\ref{s5} the experimental and theoretical results for the
neutron--capture cross section of $^{36}$S
are compared and discussed.
\section{Fast Cyclic Activation Technique}\label{s2}
The measurements have been carried out at the Karlsruhe pulsed 3.75 Van de
Graaff accelerator.
A common activation measurement is subdivided into two parts: (1) the
irradiation of the sample,
(2) the counting of the induced activity~\cite{bee80}.
The cyclic activation method is the repetition of the irradiation
and activity counting procedure many times to gain statistics. Especially
for nuclei with half lives of only minutes or seconds, in our case
$^{37}$S with a half life of 5.05 minutes,
a large number of irradiation and counting cycles is needed.
The time constants for each cycle which are chosen shorter than the
fluctuations of the neutron beam and comparable or shorter than the decay
rate $\lambda$ of the measured isotope
are, the irradiation time, $t_{\rm b}$, the counting time, $t_{\rm c}$, the
waiting time, $t_{\rm w}$ (the time to switch from the irradiation to the
counting phase), and the total time, T=$t_{\rm b}$+$t_{\rm w}$+$t_{\rm
c}$+$t'_{\rm w}$
($t'_{\rm w}$ the time to switch from the counting to the irradiation phase).
In the actual $^{36}$S measurements the runs were partly carried out with
$t_{\rm b}$=19.58\,s, $t_{\rm c}$=19.24\,s, T=40\,s, partly with $t_{\rm
b}$=49.58\,s,
$t_{\rm c}$=49.24\,s, T=100\,s.
The waiting time is in both cases $t_{\rm w}$=0.42\,s.

The accumulated number of counts from a total of n cycles,
$C=\sum_{i=1}^n C_i$, where $C_i$,
the counts after the i--th cycle, are calculated for a chosen irradiation
time, $t_b$, which is short
enough compared with the fluctuations of the neutron flux with~\cite{bee94}
\begin{equation}
\label{eq1}
C  = \epsilon_{\gamma}K_{\gamma}f_{\gamma}\lambda^{-1}[1-\exp(-\lambda
t_{\rm c})]
\exp(-\lambda t_{\rm w})
 \frac{1-\exp(-\lambda t_{\rm b})}{1-\exp(-\lambda T)} N \sigma
{[1-f_{\rm b} \exp(-\lambda T)]}
 \sum_{i=1}^n \Phi_i
 \end{equation}
with
\begin{displaymath}
f_{\rm b}  =  \frac{\sum_{i=1}^n \Phi_i \exp[-(n-i)\lambda T}{
\sum_{i=1}^n \Phi_i} \quad .
\end{displaymath}
The following additional quantities have been defined:
$\epsilon_\gamma$: Ge--efficiency, $K_\gamma$:
$\gamma$--ray absorption, $f_\gamma$: $\gamma$--ray intensity per decay,
$N$: the number of target
nuclei, $\sigma$: the capture cross section, $\Phi_i$: the neutron flux
in the i--th cycle. The
quantity $f_{\rm b}$ is calculated from the registered flux history of a $^6$Li
glass monitor.

The efficiency determination of the 35\,\% HPGe--detector (2\,keV resolution
at 1.332\,MeV) has been reported elsewhere~\cite{bee92}. The $\gamma$--ray
absorption was calculated using tables published by Storm and
Israel~\cite{sto70} and Veigele~\cite{vei73}.
The half--lives and the $\gamma$--ray intensities per decay of $^{37}$S and
$^{198}$Au are given in Table~\ref{tt1}.

The activities of nuclides with half lives of several hours
to days, i.e.~the activity of $^{198}$Au, can be also counted after
the end of the cyclic activation consisting of $n$ cycles using
\begin{eqnarray}
\label{eq2}
C_n=\epsilon_\gamma K_\gamma f_\gamma \lambda^{-1} [1-\exp(-\lambda T_{\rm M})]
\exp(-\lambda T_{\rm W})
[1-\exp(-\lambda t_{\rm b})] N \sigma f_{\rm b} \sum_{i=1}^n \Phi_i \quad.
\end{eqnarray}
Here $T_{\rm M}$ is the measuring time of the Ge--detector and $T_{\rm W}$
the time
elapsed between the
end of cyclic activation and begin of the new data aquisition.

The Eqs.~\ref{eq1} and \ref{eq2}, respectively, contain the
unknown quantities $\sigma$
and the total neutron flux $\sum_{i=1}^n \Phi_i$. Therefore, cross section
ratios can be formed
for different isotopes exposed to the same total neutron flux. This is the
basis for the determination
of the $^{36}$S capture cross section relative to the well--known standard
$^{197}$Au capture
cross section \cite{rat88}. As the $^{36}$S sample to be investigated is
characterized by a finite thickness
it is necessary to sandwich the sample by two comparatively thin gold
foils for the determination
of the effective neutron flux at sample position. The activities of these
gold foils were counted also
individually after termination of the cyclic activation. The effective
count rate of gold was obtained from these individual rates as well as
from the accumulated gold count rate during the cyclic activation run.
Therefore, the effective neutron flux at sample position was determined
in two ways by way of the gold activation according to the Eqs.~\ref{eq1}
and \ref{eq2}.
Using Eq.~\ref{eq1} has the advantage that
saturation
effects in the gold activity for irradiations over several days are
avoided~\cite{bee94}.

\section{Experimental Arrangements and Results}\label{s3}
In Fig.~\ref{ff1} a scheme of the experimental setup is shown.
The kinematically collimated neutron beam is generated with the
$^7$Li(p,n) reaction near the reaction threshold (1.912~MeV proton
energy) with thick $^7$Li--targets (30~$\mu$m) and
corresponds to a Maxwellian neutron spectrum with a thermal neutron
energy of $kT=25$\,keV
\cite{bee94,bee80,sto70,vei73,rat88}. The neutron spectra at the
neutron energies 151,
176, and 218\,keV\,keV were generated using thin Li--targets (2.5~$\mu$m).
The required proton energy
conditions and the neutron spectra integrated over the solid angle of the
sample were determined in time--of--flight (TOF) measurements
before the actual activation runs using the accelerator in pulsed
mode (Fig.~\ref{ff2}). These neutron spectra are in good agreement with
corresponding
Monte Carlo calculations~\cite{sch92} and \cite{sch95}.
The proton beam was wobbled initiated by magnetic deflection to cover
the area of the
Li--target. The beam profile formed was studied on a quartz target.
To switch back and forth
between sample irradiation and activity counting a fast sample changer
operating with compressed
air was used. Close to the beam line where the neutrons have been
generated the Ge--detector
for activity counting, well shielded by lead and Li--loaded paraffin, has
been installed (Fig.~\ref{ff1}).
During the irradiation phase the analog to digital converter was gated
to prevent data acquisition.
The relative neutron flux was recorded continuously with a $^6$Li glass
detector. During the
activity counting phase neutron generation was interrupted by a beam
stop for the proton beam.
This is essential to reduce all prompt accelerator dependent
$\gamma$--rays. The beam stop was
installed in the beam line at the accelerator hall so that
in the activity counting periods the
experimental hall was free of prompt background radiation.
In Table \ref{tt1} the sample characteristics of $^{36}$S and $^{197}$Au
and the decay properties of the product nuclei are listed.
Table \ref{tt2} gives a survey of the sample weights and the measured
$^{36}$S capture cross sections. The sulfur was pressed to
self supporting tablets of 6 mm diameter, the gold foils on back and front
side of the sulfur sample with regard to the impinging neutrons had the
same dimensions.
At 25\,keV neutron
energy measurements were carried out with sulfur sample masses between 20
and 100~mg. No significant effect from multiple neutron scattering was
observed. A sample of the sulfur powder was also heated to 300$^o$~C. No
measurable weight loss due to absorbed water was found. In Fig.~\ref{ff3}
the accumulated
$\gamma$--ray intensity from one of the $^{36}$S activations is shown. The
$\gamma$--line is well isolated on a low level of background counts.

The following systematic uncertainties were combined by quadratic error
propagation: Au standard cross section: 1.5--3\,\%, Ge--detector efficiency:
6\,\%, $\gamma$--ray intensity per decay: 0.6\,\% for the $^{37}$S and 0.1\,\%
for the $^{198}$Au decay, divergence of neutron beam: 2--12\,\%, factor
$f_b$: 1.5\,\%, sample weight: $<$0.5\,\%, and other systematic errors: 2\,\%.

\section{Theoretical Analysis}\label{s4}

\subsection{Reaction mechanism and models}
In nuclear reactions two extreme types of reaction mechanisms can exist:
the compound--nucleus (CN) and the direct (DI) process. In the CN mechanism
the projectile merges in the target
nucleus and excites many degrees of freedom of the CN. The excitation proceeds
by way of a multistep process and therefore has a reaction time typically of
the
order $10^{-16}\,$s to $10^{-20}\,$s. After this time the CN decays into
various exit channels. The relative importance of the decay channels is
determined by the branching ratios to the final states.
In the DI process the projectile excites only a
few degrees of
freedom (e.g.~single--particle or collective). The excitation proceeds in one
single step and has a characteristic time scale of $10^{-21}\,$s to
$10^{-22}\,$s. This corresponds to the time it takes the projectile
to pass through the target nucleus; this time is much shorter than the
reaction time of CN processes.

\subsection{Folding procedure}
The folding procedure is used for calculating the
nucleon--nucleus potentials to describe the
elastic scattering data and the bound states. This method has been applied
successfully in describing many nucleon--nucleus systems.
In the folding approach the nuclear density
$\rho_{A}$ is folded with an energy and density dependent nucleon--nucleon
interaction $v_{\rm eff}$~\cite{kob84}, \cite{obe91}
\begin{equation} \label{1}
V(R) = \lambda V_{\rm F}(R) = \lambda \int
\rho_{A} (\vec{r}) v_{\rm eff}(E,\rho_A,|\vec{R}  - \vec{r}|)
d\vec{r}
\end{equation}
with $\vec{R}$ being the separation of the centers of mass of the two colliding
nuclei. The normalization factor $\lambda$ is adjusted to reproduce the
elastic scattering data the binding energies of the residual nuclei.
The folding potentials of Eq.~\ref{1} were determined with the help of the
code DFOLD~\cite{abe91}.

\subsection{Direct--capture model}
In thermonuclear scenarios the projectile energy is well below the Coulomb
and/or centrifugal barrier. Consequently,
the CN formation may be suppressed, because there are almost no
CN levels that can be populated, especially in light, magic and
far--off--stability nuclei.
In this work the theoretical cross sections are calculated
by only considering the DC--contributions.
The theoretical cross section $\sigma^{\rm th}$
is obtained from the
DC cross section $\sigma^{\rm DC}$ by~\cite{obe91}, \cite{moh93}
\begin{equation} \label{2}
\sigma^{\rm th} = \sum_{i} \: C_{i}^{2} S_{i}\sigma^{\rm DC}_{i} \quad .
\end{equation}
The sum extends over the ground state and
excited states in the final nuclei, where the spectroscopic
factors $S_{i}$ are known. The isospin Clebsch--Gordan coefficients
are given by $C_{i}$. The DC cross sections $\sigma^{\rm DC}_{i}$ are
essentially
determined by the overlap of the scattering wave function
in the entrance channel, the bound--state wave function
in the exit channel and the multipole transition--operator. The radial
dependence of the wave functions
in the DC--integral is in our case determined uniquely by the folding
potentials.

\section{Calculations and results}\label{s5}
In the folding approach the nuclear density $\rho_{A}$ for the stable nucleus
$^{36}$S was derived from the experimental charge
distribution~\cite{deV87}.
The normalization factor $\lambda$ for $^{36}$S(n,$\gamma$)$^{37}$S
of the optical potential in the entrance
channel was adjusted to fit the thermal
($^{36}$S+n)--scattering cross section of ($1.1 \pm 0.8$)\,barn~\cite{sea92}.
Even so this cross section is not determined well, we fitted
our normalization factor to reproduce 1.1\,barn.
However, applying the same fitting procedure to
the ($^{34}$S+n)--scattering cross section that is known
much better (($1.52 \pm 0.03$)\,barn~\cite{sea92}) we obtained almost
the same volume integral per nucleon in the two cases
($^{34}$S+n: 501.7\,MeV\,fm$^3$; $^{36}$S+n: 497.1\,MeV\,fm$^3$).
The
imaginary part of the optical potential is small for
the ($^{36}$S+n)--channel and can be neglected.
For the exit channels the normalization constants $\lambda$
were adjusted to the energies of the
ground and the excited states. The potentials obtained in
this way ensure the correct behavior of the wave functions in the nuclear
exterior.

The spectroscopic factors for one--nucleon stripping
of $^{37}$S were determined from the most recent
experimental $^{36}$S(d,p)$^{37}$S--data~\cite{end90}.
We also carried out shell--model calculations using a
combined sd-- and pf--shell with an effective
nucleon--nucleon interaction derived by Warburton et al.~\cite{war87}.
For these
calculations the program OXBASH~\cite{bro84} was used to calculate
the wave functions and spectroscopic factors. The calculated
spectroscopic factors are larger than those extracted
from the experimental $^{36}$S(d,p)$^{37}$S--data (see Table~\ref{t1}).
This fact might be due to the strong truncation of the model space
(only one neutron in the pf--shell for the negative parity
states). Inclusion of more particle--hole excitations
should give smaller spectroscopic factors.
The masses and Q--values for the transitions
to the different states of the residual nucleus
$^{37}$S were taken from experimental data~\cite{aud93},
\cite{end90}.
For the DC--calculations the code TEDCA~\cite{kra92} was used.

The level scheme of the relevant levels for $^{37}$S
is shown in Fig.~\ref{f1}.
The cross section for the reaction $^{36}$S(n,$\gamma$)$^{37}$S
obtained from the DC--calculation is compared with the
experimental data from the thermal to the thermonuclear
energy region in Fig.~\ref{f2}.
There are two types of E1--transitions contributing for
the transitions to the residual nucleus $^{37}$S.
The first one comes from an s--wave in the
entrance channel exciting the negative--parity states
3/2$^{-}$ and 1/2$^-$ (see Table~\ref{t1}).
These transitions give the well--known
1/v--behavior (see Fig.~\ref{f1}). The second type of E1--transition comes from
an initial p--wave and excites the positive--parity state 3/2$^{+}$
in the final nucleus. This transition
has a v--behavior and can be neglected in the relevant energy range
(see Fig.~\ref{f1} and Table~\ref{t1}). The E1--transition to the
7/2$^-$--ground state
of $^{37}$S can also be neglected, because of the higher centrifugal
barrier
necessary for the incoming d--wave (see Table~\ref{t1}). As can be seen
from Fig.~\ref{f1} this contribution effects the deviation from an
1/v--behavior of the cross section only above about 700\,keV.

The spin and parity assignments of the final states in $^{37}$S,
the Q--values  for the transitions to the different final states,
the spectroscopic factors obtained from
$^{36}$S(d,p)$^{37}$S and shell--model calculations
are shown in Table~\ref{t1}. Also in this table the calculated
cross sections for $^{36}$S(n,$\gamma$)$^{37}$S
at 25.3\,meV, 25, 151,
176, and 218\,keV\,keV using DC with the
spectroscopic factors obtained from the (d,p)--data
are compared with the experimental data.

We have determined the thermonuclear--reaction--rate factor
$N_{A}\left<\sigma v \right>$~\cite{fow67}. Since the cross
section follows an $1/v$--law up to 150\,keV we obtain a
constant reaction--rate factor
\begin{equation}\label{3}
N_{A}\left<\sigma v \right> = 2.56 \times
10^{4}\,{\rm cm}^3\,{\rm mole}^{-1}\,{\rm s}^{-1} \quad .
\end{equation}

\section{Discussion}\label{s6}
Direct--capture calculations using the folding procedure
can  excellently reproduce the non--resonant experimental data
for the capture cross section by the neutron--rich sulfur isotope
$^{36}$S in the thermal and thermonuclear
energy region. The enhancement in the region of 176\,keV
comes from resonant contributions~\cite{end90} not considered in the
DC--calculation.

DC is also the
dominant reaction mechanism for neutron capture by neutron--rich isotopes
far--off stability occurring in the $\alpha$--and r--process.
For such isotopes the Q--value and therefore the excitation energy
of the compound nucleus gets still lower,
leading to a further substantial diminution of the level density of
the compound nucleus. Thus,
the DC--contribution becomes the dominating reaction mechanism.
Nuclear--structure models are indispensable for extrapolating
reaction rates to nuclei near and far--off the region
of stability, because only a limited or no
experimental information is available in this region.
The DC cross reaction $^{36}$S(n,$\gamma$)$^{37}$S  can be considered
as a benchmark of different nuclear--structure models
(Shell Model, Relativistic
Mean Field Theory, Quasi Particle Random Phase Approximation,
Hartree Fock Bogoliubov Theory) for calculating
neutron--capture cross sections by neutron--rich nuclei
taking place in the $\alpha$-- or r--process~\cite{ohu95}.

The s--process production of $^{36}$S was recently discussed quantitatively
by Schatz
et al.~\cite{sch95} but without a reliable $^{36}$S(n,$\gamma$) cross section.
The s--process reaction network in the sulfur to
calcium region contains (n,$\gamma$), (n,p), and (n,$\alpha$) reactions.
The $^{36}$S production is mediated by the $^{36}$Cl(n,p)$^{36}$S reaction
from seed nuclei with mass numbers $<$36. But also seed nuclei $>$36 can
contribute through the $^{39}$Ar(n,$\alpha$)$^{36}$S reaction channel. Besides
its formation the destruction of $^{36}$S by the
(n,$\gamma$) reaction is important. A decrease in the
$^{36}$S(n,$\gamma$)$^{37}$S
cross section leads to a corresponding increase of the abundance formed.
As our
measured $^{36}$S(n,$\gamma$)$^{37}$S  value is by a factor of 1.8 smaller
than the
estimate of Woosley et al.~\cite{woo78} the s--process abundance production
of $^{36}$S will be enhanced by this factor.
The quantitative analysis requires also model parameters for the main and
especially
the weak s--process component. This information can be obtained from the
analysis of the s--process beyond A=56~\cite{bee95}.
\section*{Acknowledgements}
We would like to thank our technician G. Rupp and the Van de Graaff staff
members,
E.-P. Knaetsch, D. Roller, and W. Seith for their skill in preparing the
metallic Li-targets in the required quality. We are also indebted to
the Van de
Graaff staff for providing reliable beam conditions of the accelerator.
We thank the  Fonds zur F\"orderung der
wissenschaftlichen Forschung in \"Osterreich (project S7307--AST)
and the \"Ostereichische Nationalbank (project 5054)
for their support. Two of the authors (H.H.~and H.O.) express
their gratitude to M.~Wiescher for valuable and helpful discussions
during their stay at the Univ.~of Notre Dame. One of the authors (H.B.)
is grateful to Ch. Theis and S. Jaag for support in special problems of the
computer system for data acquisition and handling.

\begin{table}
\caption{\label{tt1}Sample characteristics and decay properties of the
product nuclei $^{37}$S and $^{198}$Au}
\begin{center}
\begin{tabular}{ccccccc}
Isotope&Chemical & Enrichment & Reaction & $T_{1/2}$ & $E_\gamma$ &
Intensity per decay\\
  &form   &  (\%)   &   &   &  (keV)    & (\%) \\
\hline
$^{36}$S&sulfur powder& 5.933&$^{36}$S(n,$\gamma$)$^{37}$S&5.05 min& 3103 &
94.0$\pm$0.6 \\
$^{197}$Au&metallic& 100&$^{197}$Au(n,$\gamma$)$^{198}$Au&2.69 d& 412&
95.50$\pm$0.096\\
\end{tabular}
\end{center}
\end{table}

\begin{table}
\unitlength1cm
\caption{\label{tt2}Sample weights and experimental $^{36}$S
capture cross sections}
\begin{center}
\begin{tabular}{dddddddd}
Mean neutron  & Mass of Au &  Mass of  & Mass of Au &
Irradiation &
$\sigma$ &\multicolumn{2}{c}{Uncertainty}\\
energy& front side& sulfur& back side& time  & ($\mu$barn) &
statistical&total\\
(keV)   &   (mg)  &  (mg) & (mg)  & (d) &  &(\%)&(\%)\\
\hline
25    &  16.80   &  100.32& 16.84&   3.54  &  192&1.2&11.3\\
      &  16.61   &   95.32& 16.64&   1.70  &  195&1.1&10.5\\
      &  16.57   &   50.40& 16.51&   1.16  &  169&2.8&8.3\\
      &  16.57   &   50.40& 16.51&   1.70  &  168&2.3&8.1\\
      &  15.71   &  148.89& 15.73&   0.58  &  191&1.5&13.9\\
      &   8.86   &   50.11&  8.80&   2.70  &  178&1.0&8.3\\
      &  17.15   &   21.18& 17.22&   5.10  &  190&1.6&7.5\\
\cline{6-8}
\multicolumn{5}{r}{Average}&       187$\pm$14&&\\
\hline
$151 \pm 15$   & 16.38   &   94.96& 16.71&   3.70  &  85&3.3&9.5\\
       & 16.46   &  150.19& 16.33&   6.90  &  80&1.8&11.5\\
\cline{6-8}
\multicolumn{5}{r}{Average}&        81$\pm$7&&\\
\hline
$176 \pm 20$  &  17.35   &   94.94& 17.34&   2.40  &   114&7.2&11.4\\
      &  17.35   &   94.94& 17.34&   7.80  &   128&3.2&9.4\\
\cline{6-8}
\multicolumn{5}{r}{Average}&        125$\pm$11&&\\
\hline
$218 \pm 23$  &  17.39   &  150.00& 17.39&   3.20  & 87&6.3&11.2\\
      &  17.40   &  149.85& 17.38&   6.80  & 74&4.7&11.7\\
\cline{6-8}
\multicolumn{5}{r}{Average}&        78$\pm$7&&\\
\end{tabular}
\end{center}
\end{table}

\begin{table}
\caption[36S]{\label{t1} Final states, Q--values, transitions and spectroscopic
factors for the states of $^{37}$S
obtained from shell--model calculations and
$^{36}$S(d,p)$^{37}$S. Cross sections for $^{36}$S(n,$\gamma$)$^{37}$S
at 25.3\,meV, 25\,keV, 151\,keV, 176\,keV, and 218\,keV using DC
with the experimental data.}
\begin{center}
\begin{tabular}{cccccrrrrr}
Final & Q--value & Transition & \multicolumn{2}{c}{Spectroscopic factor} &
\multicolumn{5}{c}{Cross section} \\
state & [MeV] & & shell & (d,p)--reaction &
25.3\,meV & 25\,keV & 151\,keV & 176\,keV & 218\,keV \\
&&& model & \cite{end90} &
[mbarn] & [$\mu$barn] & [$\mu$barn] & [$\mu$barn] & [$\mu$barn] \\
\hline
$\frac{7}{2}^-$ & 4.303 & d $\rightarrow$ f & 0.91 & 0.50 &
0.0 & 0.0 & 0.1 & 0.2 & 0.2 \\
$\frac{3}{2}^-$ & 3.657 & s $\rightarrow$ p & 0.86 & 0.55 &
157.1 & 158.0 & 64.3 & 59.5 & 53.5 \\
$\frac{3}{2}^+$ & 2.906 & p $\rightarrow$ d & 0.07 & 0.03 &
0.0 & 0.0 & 0.0 & 0.0 & 0.0 \\
$\frac{3}{2}^-$ & 2.312 & s $\rightarrow$ p & 0.12 & 0.03 &
5.3 & 5.3 & 2.1 & 2.0 & 1.8 \\
$\frac{1}{2}^-$ & 1.666 & s $\rightarrow$ p & 0.97 & 0.47 &
28.3 & 28.5 & 11.6 & 10.8 & 9.7 \\
\cline{6-10}
\multicolumn{5}{r}{Total cross section: DC} &
190.7 & 191.8 & 78.1 & 72.5 & 65.2 \\
\cline{6-10}
\multicolumn{5}{r}{Total cross section: experiment} &
$150 \pm 30$$^{1}$
&  $187 \pm 14$ & $81 \pm 7$ & $125 \pm 11$ & $78 \pm 7$ \\
\end{tabular}
\end{center}
{\footnotesize $^1$ Ref.~\cite{sea92}}
\end{table}

\begin{figure}[t]
\unitlength1cm
%\begin{center}
%\begin{picture}(12,12)
%\put(2,0){\dashbox{0.2}(10,11)[b]{}}
%\end{picture}
%\end{center}
\caption{\label{ff1} Scheme of experimental setup}
\end{figure}

\begin{figure}[t]
\caption{\label{ff2} Neutron spectra with mean neutron energies of
151(top), 176(middle), and 218\,keV(bottom) at sample position.
Using the proton energies
and the metallic Li-targets (2.5~$\mu$m thickness) of the actual
activation measurements they were calculated
from TOF spectra measured under different angles with respect to the beam axis.
}
\end{figure}

\begin{figure}[t]
\caption{\label{ff3} Accumulated intensity of the 3103\,keV
$^{37}$S $\gamma$--line
from the activation with a 100.32~mg sulfur sample}
\end{figure}

\begin{figure}[t]
\caption{\label{f1} Level scheme of $^{37}$S}
\end{figure}
\begin{figure}[t]
\caption{\label{f2} Comparison of the DC cross section for
$^{36}$S(n,$\gamma$)$^{37}$S
with the experimental data from thermal to thermonuclear projectile energies.
The DC contributions for the different transitions to the final states of
$^{37}$S as well as the sum of all transitions (solid curve) is shown. The
experimental data at the thermal energy have been taken from Ref. [35].
}
\end{figure}

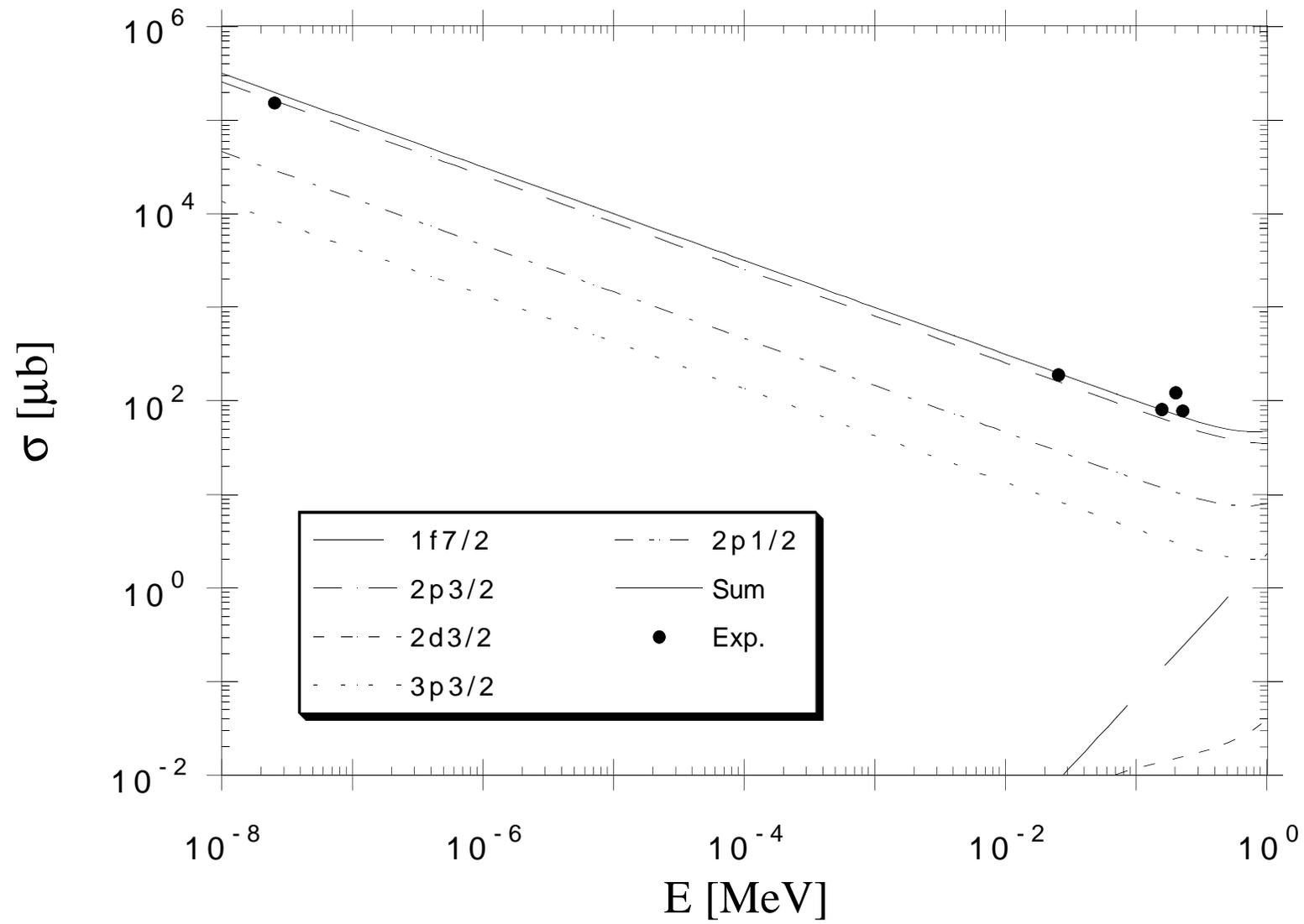
\end{document}